\documentclass[aps,prb,groupedaddress,showkeys,preprintnumbers,preprint]{revtex4}

\usepackage{graphicx}
\usepackage{color}
\usepackage[ansinew]{inputenc}
\usepackage{amsmath}
\usepackage{amsfonts}
\usepackage{amssymb}
\usepackage{dcolumn}
\usepackage{float}

\begin{document}
\title{\ \\{}Random sequential adsorption model of damage and\\{}crack
    accumulation: Exact one-dimensional results\\{}\ }

  \author{{\bf Oleksandr\ Gromenko},\ {\bf Vladimir\ Privman}\ and\ {\bf M.\ L.\ Glasser}}
  \affiliation{Department~of~Physics,~Clarkson~University,\\{}Potsdam,~New~York~13699--5721,~USA{}\\{}\ {}\\{}\ }

  \begin{abstract} 
    The random sequential adsorption (RSA) model is modified to describe damage
and crack accumulation. The exclusion for object deposition (for
damaged region formation) is not for the whole object, as in the standard
RSA, but only for the initial point (or higher-dimensional defect) from
which the damaged region or crack initiates. The one-dimensional 
variant of the model is solved exactly.\\

\noindent{\bf Posted as e-print 0712.3567 at www.arxiv.org} \\

\noindent{\bf Journal of Computational and Theoretical Nanoscience, in print (2008)}\\
 \end{abstract}


\maketitle

\section{Introduction}

Random sequential adsorption (RSA) has found many applications,\cite{Evans,PrBr} 
notably in irreversible surface deposition.\cite{PrJA,PrCS} In this article, we 
consider a variant of the RSA model motivated by study of damage accumulation, e.g., 
multicracking. The first adaptation of the RSA model in mechanical-engineering
literature to analyze
fiber fragmentation in a unidirectional composite,\cite{Curtin} was followed
by   defect modeling in thin coatings.\cite{Calka,Zhao,Mezin,Hui} In parallel,
RSA-type models were also utilized in the statistical mechanics literature to
explore approaches to damage development in several situations.\cite{Kr2,Kr1,Bradley}

Formation of a crack relieves stress in the nearby material, thus
creating a region along the crack, possibly exclusive of the crack 
tip/edge area, in which new cracks cannot be initiated.
Cracks formed in a thin coating can
be approximated as a collection of one-dimensional (1D) ``rods''
(segments), subject to the no-overlap restriction as they form
sequentially, similar to the traditional RSA. Experiments on
multiple cracking in metal\cite{Ramsey,Shieu,Hu} and 
nonmetal\cite{Agrawal,Watkins,Delannay} coatings have been  modeled in this way. 
Furthermore, experiments\cite{Fraser,Netravali,Wagner,Rao} on load-induced damage
in uniaxial fiber materials have also been interpreted in terms of
RSA-type models that involve mathematics\cite{Curtin,Hui} similar to that
encountered in the present article.

Generally, in materials subject to stress, cracks are initiated from
static microscopic defects.\cite{Lawn,Broberg} In the standard
formulation of the RSA model, newly formed (or deposited on
surfaces) objects cannot overlap  objects  formed  earlier. Here we
consider a variant of the RSA model that is arguably a better
representation for damage accumulation in materials. Indeed, since
the formed cracks relieve stress, except perhaps at their 
edges (as long as they grow), then new cracks cannot be initiated at
or near the existing cracks. Therefore, as will be described later,
the RSA model has to be modified to allow for ``exclusion'' of only
the initial region of the crack. We do allow new cracks to be initiated 
close to the tips of the earlier formed cracks.

Despite  the seeming simplicity of the RSA model, exact results
are usually available only in one dimension\cite{PrivmanRSA,Gonzales} 
and in other special lattice geometries.\cite{Evans}
In this article, we consider a 1D model of deposition of segments of
length $\ell$ on a line, and we allow exclusion to apply only to the
head points of the segments. We argue that this model describes the
accumulation of cracks which can reach lengths up to $\ell$ in 1D
geometries.

In Section\ \ref{model}, we describe the model and then in Section\
\ref{segments}, carry out an exact calculation of the open-interval
probability, which allows us to evaluate the density of cracks as a
function of time, $t$. In Section\ \ref{gaps}, we analyze the
probability distribution of gaps, exactly and asymptotically for
large times.

\section{RSA of segments with single-end exclusion}\label{model}

\begin{figure}[t]
\begin{center}
\includegraphics[width=16cm]{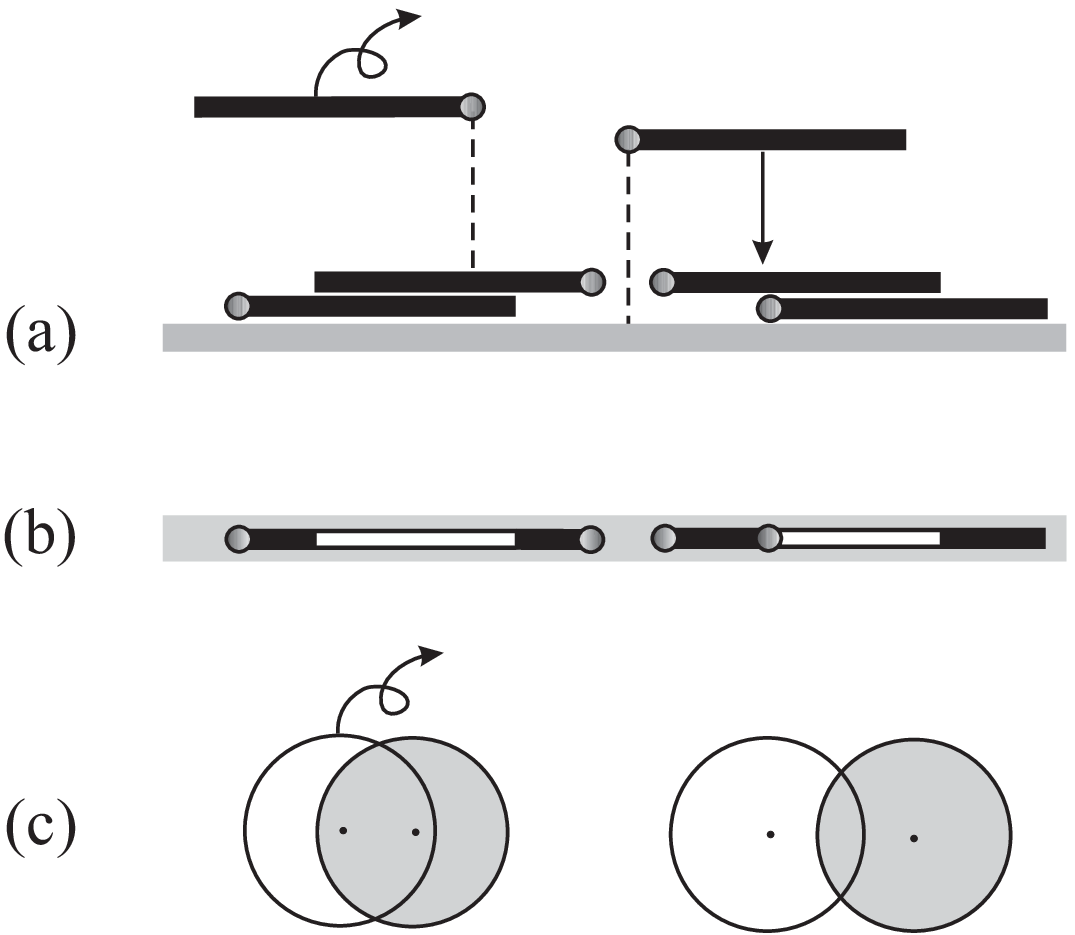}
\caption{Definition of the model: (a) Examples of possible
configurations of sequential deposition of segments with head-point
exclusion on a line. The curly arrow marks an arriving segment the
deposition attempt of which will be rejected. (b) Formation of
cracks that maps onto the deposition problem corresponding to the
deposited pairs of segments in panel (a) above. The white area
denotes the
parts of the later-formed cracks that overlap already formed
cracks. (c) A higher-dimensional example: Here circular damage
areas are assumed to spread from center points. New damage area
cannot be initiated from a center-point that overlaps earlier
damaged areas. A configuration that will not
be realized is marked by a curly arrow. As further explained in the text,
this model illustrates that the point-exclusion RSA can be nearly
identical to the ordinary RSA (here, of deposition of half-radius circles) 
especially when the objects and the exclusion point location are 
symmetrical.}\label{F1}
\end{center}
\end{figure}

Our  model consists of the deposition of segments of length $\ell$ on a
linear substrate, as illustrated in Figure\ \ref{F1}(a). The
segments are transported to the substrate at the rate $R$ (number of
deposition attempts per unit length per unit time). Each
segment has a ``head'' point, and the segment orientation
is otherwise random: right-to-left or left-to-right segments arrive
at the rates $R/2$ per unit length. Unlike the
ordinary RSA,  exclusion applies only to the adsorption of head points,
which can be deposited only in areas that are not covered by the
previously deposited segments. The lower pairs of segments in
Figure\ \ref{F1}(a) show two such possible deposition sequences.
However, the segments with head points arriving above the covered
areas are rejected (their deposition attempts fail and they are
transported away from the substrate). One such example is illustrated in Figure\ \ref{F1}(a).

Figure\ \ref{F1}(b) illustrates the equivalence of the deposition
model to crack formation. The head points of the segments correspond
to the initiation points of the cracks in a 1D structure, e.g.,
fiber.\cite{Curtin,Calka,Zhao,Mezin,Hui} For tractability (otherwise 
our model would not be exactly solvable by the utilized technique), we 
assume that the cracks grow
up to a fixed length $\ell$, except when they run into the 
cracks formed earlier, as shown in the figure for two configurations, each
corresponding to the first two segments deposited in the
corresponding locations in Figure\ \ref{F1}(a). Furthermore, we consider the
situation where the cracks grow instantaneously once initiated, as compared
to the time scales of new crack initiation. 

Since there is a zone near a crack with reduced stress (except perhaps
near the tip/edge), new cracks
cannot initiate nearby. In fact, in real materials
higher-dimensional damage configurations should be considered. They
can originate not only from point-like defects and not all be
identical in size and shape.\cite{Lawn,Broberg} A simple
illustration is given in Figure\ \ref{F1}(c). Here circular damage
regions originate from random centers in a plane. Additional damage
can be initiated only from a center-point located in the previously
undamaged areas. Figure\ \ref{F1}(c) illustrates that the present 
models can be quite close to the ordinary RSA models. Indeed, unless
the exclusion point is taken off-center or the objects are made less
symmetrical, the model in the figure is nearly identical to the ordinary
RSA of half-diameter circles (with no-overlap exclusion): the only
difference is in the way the ``covered'' area is counted, whereas the
center deposition kinetics is identical for both models.

We note that in the crack-formation nomenclature the parameter $R$
corresponds to the rate (per unit time) of crack initiation per unit length 
of the 1D structure. Further discussion (and literature citation) is available\cite{DP} on 
how such statistical-mechanical parameters relate to the actual system
properties. In this work we consider the simplest model, the main virtue 
of which is the possibility of deriving 
analytically exact or asymptotic results. More complicated models, which can 
be realistically compared with experimental data, would require large-scale numerical
simulations and will be the subject of future work.
 
The quantities of interest for the model just defined are as
follows. It is interesting to calculate the fraction of the total
length that is undamaged, $F(t)$, which is expected to decrease
monotonically from $F(0)=1$ to $F(\infty) =0$. A second quantity of
interest is the density of deposited segment heads (the density of
cracks initiated), $N(t)$. Obviously, \hbox{$N(0)=0$} because, for
simplicity, we assume that initially the system is undamaged.
However, we will establish that this quantity actually diverges
logarithmically as $t \to \infty$. The calculation of $F(t)$ and
$N(t)$ will be carried out exactly in the next section.

Another quantity of interest is the density (per unit length),
$Q(x,t)dx$, of gaps (between the damaged areas) of sizes from $x$ to
$x+dx$. The quantity, $Q(x,t)$, gives us the size-distribution of
undamaged regions in the 1D material sample. The total density
(number per unit length) of gaps between the damaged regions is then
\begin{equation}\label{GGG}
G(t)=\int_0^\infty Q(x,t) dx \, .
\end{equation}
Note that $Q(x,0)=0$ and, obviously,
\begin{equation}\label{GGG2}
G\left(t \ll (R\ell)^{-1}\right) \simeq R  t \, .
\end{equation}

In addition to exact results, it is of interest to
explore the small-gap statistics leading to the
derivation of the asymptotic results for $Q(x,t)$
in the large-time regime, when the total number of
gaps has reached the
saturation value $G(\infty)$. This matter will be
further
elucidated in Section\ \ref{gaps}, where the
results for the gap density are presented.

\section{Empty-interval probability}\label{segments}

Exact results for the present model can be derived for the
empty-interval probability, $P(x,t)$. This quantity represents the
probability that a randomly chosen interval of length $x$ is fully
undamaged, i.e. open for deposition of segment-heads. Since such an
interval could be a part of a longer empty region,  $P(x,t)$ is {\it
not\/} proportional to the gap density $Q(x,t)dx$ introduced
earlier. These quantities will be related in Section\ \ref{gaps}.
However, one can easily convince oneself that the fraction of the
total length that is undamaged is given by
\begin{equation}
F(t)= P(0,t)\, .
\end{equation}

The probability $P(x,t)$ satisfies the equation
\begin{equation}
\label{Prob} \frac{\partial P(x,t)}{\partial t}=-R\,x\,
P(x,t)-R\!\int\limits^{x+\ell}_{x}\!dy\,P(y,t)\, ,
\end{equation}
where the first term on the right-hand side accounts for depositions where a head-point
lands in the interval $x$. The second term on the right-hand side corresponds to
deposition of segments with the head-point landing outside the interval
$x$, which requires that a larger interval, of length $y$ (up to
$x+\ell$), be open for deposition.

For the initial condition $P(x,0)=1$, the $x$ dependence in (\ref{Prob}) 
can be eliminated by the Anzatz
\begin {equation}
\label{anzatz} P(x,t)=\exp(-Rxt)\,F(t)\, .
\end{equation}
Substituting (\ref{anzatz}) in (\ref{Prob}), one obtains a
differential equation for $F(t)$ with the initial condition
$F(0)=1$. The result is
\begin{equation}
 \label{Solution}
P(x,t)=\exp\!\Big[-Rxt +
\int\limits^{t}_{0}\frac{\exp(-R\ell u)-1}{u}du\Big]\, .
\end{equation}

For the fraction of the undamaged area, we have
\begin{equation}
\label{expres_P}
F(t) = \frac{e^{-\gamma}}{R\ell t}e^{- E_{1}\left(R\ell t\right)}\, ,
\end{equation}
where $E_{1}(\tau) =
\int\limits^{\infty}_{\tau}\frac{e^{-\omega}}{\omega}d\omega$ is a standard exponential integral function, and
$\gamma=0.57721566490\dots$ is the Euler gamma constant.\cite{Gamma} 
For large times, the fraction of the undamaged
(available for deposition) length vanishes according to
\begin{equation}\label{FTFT}
F\left(t \gg (R\ell)^{-1}\right) = \frac{e^{-\gamma}}{R\ell t}\Big ( 1-\frac{e^{-R\ell t}}{R\ell t}+\cdots \Big ) \approx \frac{e^{-\gamma}}{R\ell t}\, ,
\end{equation}
where from now on the sign $\approx $ will denote the large-time asymptotic behavior.

While the total undamaged length shrinks to zero at large times,
the total density of the deposited segments (the total density of
cracks initiated), $N(t)$, actually diverges logarithmically, 
\begin{equation}\label{NNN}
N\left(t \gg (R\ell)^{-1}\right) \approx \frac{e^{-\gamma}}{\ell}\ln(R\ell t)\, .
\end{equation}
This behavior, characteristic of the present model, is opposed to ordinary 1D RSA.
Indeed,
\begin{equation}
 N(t)=R\int\limits^{t}_{0}F(v)dv=R\int\limits^{t}_{0}\exp\left[\int\limits^
 {v}_{0}\frac{e^{-R\ell u}-1}{u}du\right]dv\, .
\end{equation}
This can be also written
\begin{equation}
 N(t)=\frac{e^{-\gamma}}{\ell}\Big[ e^{-E_{1}(R\ell t)}\ln(R\ell t)  - \! \int\limits^{R\ell t}_{0}
 \! du\frac{\ln u}{u}e^{-u}e^{-E_{1}(u)}\Big]\, ,
\end{equation}
where the first term dominates the divergence (\ref{NNN}).

\section{The Gap Density Distribution}\label{gaps}

In order to relate the gap density distribution $Q(x,t)$ to the
empty interval probability $P(x,t)$, let us discretize the distances
in the problem in steps $\Delta x$, where ultimately, $\Delta x \to
0$. Specifically, the linear substrate will become a lattice of
spacing $\Delta x$, and all the other lengths, including $x$ and
$\ell$, will be assumed multiples of $\Delta x$.

Let us denote by $\Omega (x,t)$ the probability that a randomly
chosen interval of length $x=n\Delta x$ has its rightmost lattice
site blocked, while all the other, $n-1$,  lattice sites are empty.
We can write a relation between probabilities,
\begin{equation}
P(x - \Delta x,t)=P(x,t) + \Omega(x,t)\, ,
\end{equation}
which, in the limit of small $\Delta x$, gives
\begin{equation}
\Omega(x,t)=-{\partial P(x,t) \over \partial x} \Delta x \, .
\end{equation}

Let us now define the probability $\Phi (x,t)$ that a randomly
chosen interval of length $x=n\Delta x$ has both its rightmost
lattice site and its leftmost lattice site blocked, while all the
other, $n-2$,  lattice sites are empty. The probability relation is
then
\begin{equation}
\Omega(x - \Delta x,t)=\Omega(x,t) + \Phi(x,t)\, ,
\end{equation}
from which we get, for small $\Delta x$,
\begin{equation}
\Phi(x,t)={\partial^2 P(x,t) \over \partial x^2} (\Delta x)^2 \, .
\end{equation}

\begin{figure}[t]
\begin{center}
\includegraphics[width=16cm]{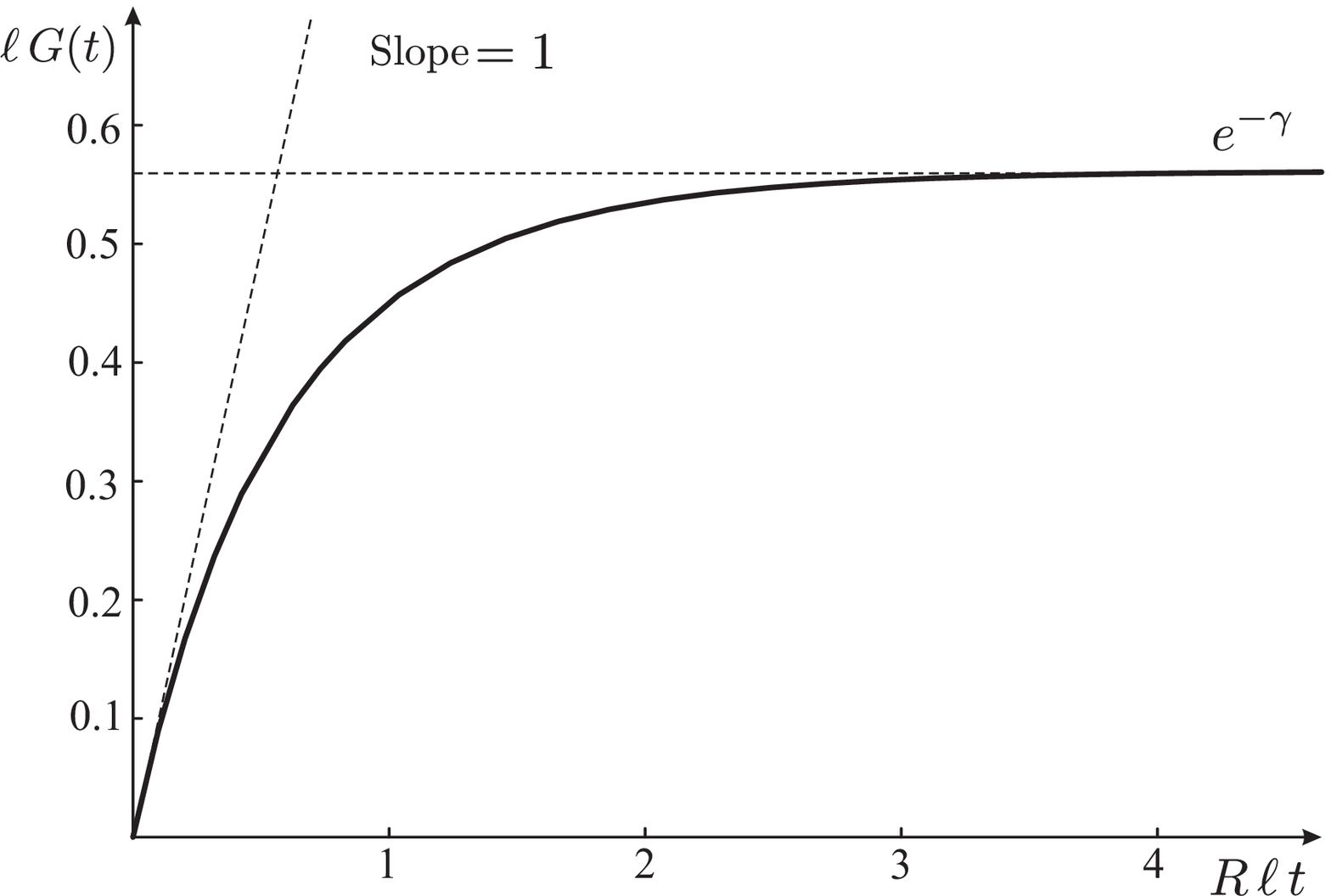}
\caption{The behavior of the density of gaps as a function of time.
The asymptotic large-time regime is realized to a good approximation
for $t > t_0$, where $t_0 \simeq 3(R\ell)^{-1}$.}\label{F2}
\end{center}
\end{figure}

Since the density (per unit length) of fixed-length intervals on the
lattice is $1/\Delta x$, for the density of gaps of size $x$ we get
\begin{equation}
Q(x,t)\Delta x=\Phi(x,t)/\Delta x={\partial^2 P(x,t) \over \partial x^2} \Delta x \, .
\end{equation}
Finally, we get
\begin{equation}\label{FMFN}
Q(x,t)={\partial^2 P(x,t) \over \partial x^2} = ( Rt/\ell )e^{- Rxt-E_{1}\left(R\ell t\right)-\gamma} \, .
\end{equation}
The total density of gaps per unit length, cf.\ (\ref{GGG}), is thus
\begin{equation}
G(t)=e^{-E_{1}\left(R\ell t\right)-\gamma} \big / \ell\, .
\end{equation}
This function is shown in Figure\ \ref{F2}.

Since exact solutions are not possible in most geometries,
considerations that can yield asymptotic large-time behavior play an
important role in the study of RSA models. We point out that in the
opposite,  short-time limit, when  deposition events are initially
not correlated, a low-density approximation leads to relations of
the type (\ref{GGG2}). The large-time behavior, however, is less
straightforward. It can be analyzed to various degrees, depending on
the model details, by considering the dynamics of gaps small enough
such that their evolution can be described approximately in term of
their size and shape. Specifcially, for ordinary RSA one 
considers\cite{Pomeau,Swendsen,Wang} the regime of late times when most of
the remaining gaps are small enough to accommodate at most one
depositing object. One further assumes that the distribution of
these gaps at some large $t_0$, from which time on the
small-gap-dominated behavior sets in, is smooth with respect to
size, and their shape distribution is also known (since we know the
geometry of the deposited objects surrounding the gaps) and is
similarly smooth. The kinetics of these gaps, blocked by the
arriving depositing object, is then evaluated\cite{Pomeau,Wang} in
order to obtain the evolution of the gap size and shape
distributions for $t>t_0$, from which  other deposit properties can
be found.

In our case, it is natural to consider the large-time behavior of
the gap distribution under the assumption that it is dominated by
gaps not larger than the segment size $\ell$. Indeed, such gaps
cannot be fragmented by segment deposition: they can only be
shortened. Therefore, the asymptotic regime is for times $t>t_0$,
see Figure\ \ref{F2}, late enough for the function $G(t)$ to have
reached its constant value. However, other quantities, including the
gap distribution $Q(x,t)$, with $x \le \ell$, still vary with time.
The equation that describes this variation is
\begin{equation}
\label{QF} \frac{\partial Q(x,t)}{\partial t}\approx - R\,x\,Q(x,t) +
R\!\int\limits^{\ell(\to \infty)}_{x}\!Q(y,t)dy\, .
\end{equation}
Here, not only is the equation  already approximate, but we seek the
asymptotic large-time solution in the regime dominated by small
gaps, for which we can further replace $\ell \to \infty$ in the
upper limit of the integration, as shown, because $\ell$ actually
plays no role in the evolution of small gaps (most of the length of
the deposited segments covers already blocked area and is not active
in further blocking).

The solution of the resulting equation is
\begin{equation}
Q(x,t)\approx qte^{-Rxt}\, ,
\end{equation}
where $q$ is a constant that cannot be determined from the
large-time analysis. It can be found by assuming properties for the
distribution at a reference time $t_0$, as mentioned for the
standard RSA problem. Alternatively, we can calculate (or
numerically estimate) some less detailed property related to the gap
density distribution, for instance the fraction of the total length
that remains open for deposition
\begin{equation}
\label{conts} F(t)= \int\limits^{\infty}_{0}xQ(x,t)dx \approx {q
\over R^2t } \, .
\end{equation}
Here we just use the exact large-time expression (\ref{FTFT}) to get
\begin{equation}
q = Re^{-\gamma}\big / \ell  \, ,
\end{equation}
which, finally, gives the asymptotic form of the gap density distribution,
\begin{equation}
Q(x,t)\approx Rte^{-Rxt-\gamma}\big / \ell \, ,
\end{equation}
consistent with the exact result (\ref{FMFN}).

In summary, we have studied an exactly solvable RSA-type model
motivated by the problem of accumulation of cracks in a 1D system. The
exclusion condition applies only to the head-points of the segments,
deposition of which models crack formation. Study of more realistic
systems in higher dimensions will require numerical simulations. We
point out that even in the simplest 1D model considered here, some
interesting quantities are not amenable to exact solution. For
example, we can define the probability, $B(x,t)dx$, of the blocked
intervals of length from $x (\ge \ell)$ to $x + dx$, similarly to
the distribution of the unblocked gaps, $Q(x,t)dx$. The quantity
$B(x,t)$ is of particular interest because it is expected to have a
delta-function singularity at $x = \ell$, corresponding to a final
(initially increasing, but eventually decaying as $t \to \infty$)
density of isolated deposited segments forming blocked-intervals of
length exactly $\ell$. We were not able to calculate $B(x,t)$
exactly. Finally, we explored the large-time small-gap-dominated
asymptotic behavior along the lines of the derivation appropriate to 
standard RSA models.

The authors thank Prof.\ A.\ Cadilhe for instructive discussions,
and acknowledge support of their research by the ARO under grant
W911NF-05-1-0339 and by the NSF under grants DMR-0509104 and
CCF-0726698.

{

\end{document}